\documentclass[fleqn]{article}
\usepackage{longtable}
\usepackage{graphicx}

\begin {document}
\begin{flushleft}
{\LARGE
{\bf Comment on ``Energy levels and radiative rates for Ne-like ions from Cu to Ga" by  N.~Singh and S.~Aggarwal [Pramana -- J. Phys.   89 (2017) 79]}
}\\

\vspace{1.5 cm}

{\bf {Kanti  M  ~Aggarwal}}\\ 

\vspace*{1.0cm}

Astrophysics Research Centre, School of Mathematics and Physics, Queen's University Belfast, \\Belfast BT7 1NN, Northern Ireland, UK\\ 
\vspace*{0.5 cm} 

e-mail: K.Aggarwal@qub.ac.uk \\

\vspace*{0.20cm}


\vspace*{1.0 cm}


\vspace*{1.0 cm}
\vspace*{1.0 cm}

\hrule

\vspace{0.5 cm}

\end{flushleft}

\clearpage


\begin{abstract}

Recently, N.~Singh and S.~Aggarwal [Pramana -- J. Phys.   89 (2017) 79] have reported  energies and lifetimes for 127  levels of three Ne-like ions, namely Cu~XX, Zn~XXI and Ga~XXII. For the calculations they have adopted two independent atomic structure codes, i.e. GRASP and FAC, and have concluded that both codes give comparable energies. However, we find that the differences between the two sets of energies are up to  1.5~Ryd (over 1.6$\times$10$^5$ wavenumbers) for many levels, and for all three ions. In the absence of other available theoretical or experimental data,  it becomes difficult to know which set of energies is more accurate. Through our calculations with the same code we demonstrate that their listings from the FAC calculations  are incorrect. A few more anomalies noted in their tabulated results are also highlighted.  

\end{abstract}

\clearpage

\section{Introduction}

In a recent paper, Singh and Aggarwal \cite{sa} have reported results for  energy levels, radiative rates (A-values), and lifetimes ($\tau$) among 127 levels of three Ne-like ions, namely Cu~XX, Zn~XXI and Ga~XXII. These levels belong to the 2s$^2$2p$^6$, 2s$^2$2p$^5$$3\ell$, 2s2p$^6$$3\ell$,  2s$^2$2p$^5$4$\ell$,  2s2p$^6$$4\ell$ and 2s$^2$2p$^5$5$\ell$ ($\ell \le$ 3) configurations. For the calculations, they have adopted  two independent atomic structure codes, i.e.  the general-purpose relativistic atomic structure package (GRASP)  and the flexible atomic code (FAC), which are  available at the websites \\{\tt http://amdpp.phys.strath.ac.uk/UK\_APAP/codes.html} and {\tt https://www-amdis.iaea.org/FAC/}, respectively.  It was done for the assessment of accuracy, particularly for the energy levels, because prior similar data, experimental or theoretical,  for most of the levels for these three ions do not exist. Therefore, based on the two calculations they concluded that both codes provide `comparable energies'. However, we notice that for many levels of all three ions the differences between the two sets of energies are significant, i.e. up to 1.5~Ryd -- see for example, levels 111--127 of Cu~XX, 112--119 of Zn~XXI and 111-116 of Ga~XXII in their tables 1--3.  In our long experience for a wide range of ions we have not noticed such large differences between calculations with these two codes, particularly when the same level of CI (configuration interaction) has been used. Additionally, in the absence of any other similar data, it is difficult to know which set of energies is more accurate. Therefore, we have performed our own calculations with these two codes and note that while their energies with the GRASP are correct, with FAC are not. 

\section {Energy levels}

In both calculations (with GRASP and FAC) Singh and Aggarwal  \cite{sa} have  included CI  among 51 configurations, namely 2s$^2$2p$^6$, 2s$^2$2p$^5$$n\ell$ (3 $\le n \le$ 7, but $\ell \le$ 3), 2s2p$^6$$n\ell$ (3 $\le n \le$ 7, but  $\ell \le$ 3), 2s$^2$2p$^4$3$\ell$3$\ell'$  and 2s$^2$2p$^4$3$\ell$3$\ell'$. These configurations generate 1016 levels in total, but the results have been reported for only among 127, as noted in Section~1.  Since both codes are fully relativistic and the same CI has been used in both calculations, the results obtained are {\em expected} to be comparable, and this has also been concluded by them. However, some of the energies obtained  between the two calculations differ by up to $\sim$1.5~Ryd as noted already, and therefore their conclusion is not based on the calculated data. Since such discrepancies unnecessarily confuse the users of data, we have performed our own calculations with both codes, and with the same CI. Their listed energies obtained with GRASP are found to be comparable with our own calculations, but {\em not} with FAC. We discuss these in detail below.

In tables 1--3 we list our calculated energies with FAC along with those of Singh and Aggarwal \cite{sa} for 127 levels of Cu~XX, Zn~XXI and Ga~XXII. Only their final energies, obtained with the inclusion of Breit and quantum electrodynamic effects (QED), are listed in these tables, but with both the GRASP and the FAC. Two conclusions can be easily drawn from these tables. Firstly, there are no appreciable differences between the GRASP and FAC2 (our calculations) energies for any of the ion considered here. This result is fully expected and has been noted in the past for several other ions. Secondly, and more importantly, there are significant differences between the GRASP and FAC1 energies obtained by Singh and Aggarwal. Differences in FAC1 and FAC2 energies for the lowest about 100 levels of these ions are comparatively minor, and are below 0.2~Ryd. However, for the higher excited levels the discrepancies are much larger, up to $\sim$1.5~Ryd, or equivalently over 1.6$\times$10$^5$~cm$^{-1}$ -- see for example, levels 111--127 of Ga~XXII in table~3, and in a majority of cases the results of Singh and Aggarwal are {\em lower}. We discuss below the (possible) reason for these discrepancies.

In calculations with FAC it is much easier to include a larger number of configurations and their levels, or the CSFs (configuration state functions).  For this reason, in the GRASP calculations Singh and Aggarwal \cite{sa} have ignored configurations with $\ell > $ 3, but have (perhaps) included in FAC. As a result of this the FAC calculations have been performed with 1112 CSFs, which include all possible values of $\ell$ for 4 $\le n \le$ 7 of the configurations listed above. Unfortunately, the 96 levels arising from these `additional' 12 configurations do not lie above the 1016 included in GRASP, or above the 127 listed by them, but {\em intermix} with those. In fact they intermix quite early, from about level 100 onward. In listing the FAC energies, Singh and Aggarwal have ignored this reality and have (perhaps) listed the lowest 127 energies, and hence the discrepancies.  It may be worth noting here that  their calculations with both the GRASP and FAC codes had similar discrepancies  in the past, for a range of ions,  see for example the energy levels of five Br-like ions \cite{brlike} with 38 $\le$ Z $\le$ 42 and F-like W~LXVI \cite{w66a},\cite{w66b}. 

Apart from these major discrepancies noted above, there are a few more (minor) anomalies in the tabulations provided  by Singh and Aggarwal \cite{sa}. We highlight only three, i.e. (i)   in table~1 the NIST energy for level 25 should be 79.466~Ryd, and not 7.9E+07,  (ii) the level 88 in table~1 should be $^3$F$_3^o$, and not $^3$F$_0^o$, and finally (iii) the  level 48 in table~2 should be $^3$P$_0^o$, and not $^3$F$_0^o$, as listed.  Similarly, for a few levels the FAC1 energies listed by them are non degenerate, but is not the case in our FAC2 calculations, see for example the levels 74/75 in their table~1 for Cu~XX, and 62/63, 88/89, 102/103 and 106/107 in table~3 for Ga~XXII. There are a few more anomalies in other tables as well, but we will like to particularly comment on the comparisons of energies shown in their table~4, because the discrepancies for a few levels appear to be very significant  ($\sim$2~Ryd) -- see for example, levels 3 and 5, i.e. 2s$^2$2p$^5$3s~$^1$P$_1$ and 2s$^2$2p$^5$3s~$^3$P$_1$. This is because these levels (and many more) are highly mixed and their ordering may change with the change in CI. For this reason such levels are not given an LSJ$^{\pi}$ designations in the NIST listings, but only their $J$ values.

\section{Radiative rates}

Singh and Aggarwal \cite{sa} have also listed A-values for some resonance transitions in tables~5--7, but for four types, i.e. electric and magnetic dipole (E1 and M1)  and quadrupole (E2 and M2). In table~9 they have compared their results with the earlier ones of Hibbert {\em et al} \cite{hlm} for a few E1 transitions, and have `concluded' a good agreement, whereas we find that the listed A-values of \cite{hlm} are {\em larger} by a factor of three, for all transitions and ions. This is not only contrary to their conclusion but also inconsistent with their subsequent results of $\tau$ (listed and compared in table~10) for which there are no discrepancies.  A closer look at the tables of Hibbert {\em et al} \cite{hlm} reveals that they have listed {\em weighted} A-values, i.e. $\omega$A, and for the transitions in table~9 of  Singh and Aggarwal, the statistical weight $\omega$ is exactly three, and hence the discrepancies. Similar discrepancies, and for the same reason,  were also noted earlier \cite{jqs} with their results for the transitions of Sr~XXX. 

\section{Conclusions}

In this paper we have highlighted some of the errors and discrepancies in the reported data of Singh and Aggarwal \cite{sa} for 127 levels of three Ne-like ions: Cu~XX, Zn~XXI and Ga~XXII. Particularly for energy levels, their listed results with the FAC calculations are incorrect, by up to 1.5~Ryd. Apart from this, their listed 127 levels of the 2s$^2$2p$^6$, 2s$^2$2p$^5$$3\ell$, 2s2p$^6$$3\ell$,  2s$^2$2p$^5$4$\ell$,  2s2p$^6$$4\ell$ and 2s$^2$2p$^5$5$\ell$ ($\ell \le$ 3) configurations are {\em not} the lowest in energy, because some from the neglected configurations, such as 2s$^2$2p$^5$5g, intermix. Therefore, there is scope for the improvement over their work. For the similar reason, some of the listed lifetimes may be affected because some of the missing transitions may make a contribution. Furthermore, the limited data reported in their paper for the A-values  are not of much use, because for any modelling or diagnostic application a complete set of data covering {\em all} transitions is preferred. 

With the ready availability of various atomic structure codes it has become comparatively easy to produce atomic data. However, it is still not straightforward to produce accurate and reliable data which can be applied with confidence. Most of the discrepancies, often noted in various atomic parameters, including energy levels and radiative rates, are because of the non practical assumptions made or inadequate comparisons and assessments. Many times the assessment of accuracy is based on speculations  rather than a rigorous and robust analysis. Several instances of large discrepancies in various parameters, and their possible simple resolutions, have recently been highlighted by us \cite{kma}. 

A complete set of data for energy levels, A-values and lifetimes for all three ions (Cu~XX, Zn~XXI and Ga~XXII) are reported in our recent paper \cite{nelike} and can be confidently applied in the modelling of plasmas.


\newpage

\begin{table}
\caption{Comparison of energies (in Ryd) for 127 levels of  Cu XX.} 
\begin{tabular}{rllrrrrr} \hline
Index  &     Configuration        & Level               & GRASP        &    FAC1    &  FAC2       \\
 \hline
    1  &    2s$^2$2p$^6$	  &	$^1$S$  _{0}$	&    0.0000   &   0.0000   &	   0.00000   \\
    2  &    2s$^2$2p$^5$3s	  &	$^3$P$^o_{2}$	&   70.6919   &  70.7705   &	  70.63754   \\
    3  &    2s$^2$2p$^5$3s	  &	$^1$P$^o_{1}$	&   70.8653   &  70.9535   &	  70.80313   \\
    4  &    2s$^2$2p$^5$3s	  &	$^3$P$^o_{0}$	&   72.2166   &  72.2977   &	  72.15617   \\
    5  &    2s$^2$2p$^5$3s	  &	$^3$P$^o_{1}$	&   72.3172   &  72.4053   &	  72.25147   \\
    6  &    2s$^2$2p$^5$3p	  &	$^3$S$  _{1}$	&   73.3742   &  73.4435   &	  73.32529   \\
    7  &    2s$^2$2p$^5$3p	  &	$^3$D$  _{2}$	&   73.6392   &  73.7184   &	  73.58100   \\
    8  &    2s$^2$2p$^5$3p	  &	$^3$D$  _{3}$	&   73.8841   &  73.9592   &	  73.83012   \\
    9  &    2s$^2$2p$^5$3p	  &	$^1$P$  _{1}$	&   73.9583   &  74.0387   &	  73.90212   \\
   10  &    2s$^2$2p$^5$3p	  &	$^3$P$  _{2}$	&   74.1445   &  74.2258   &	  74.08412   \\
   11  &    2s$^2$2p$^5$3p	  &	$^3$P$  _{0}$	&   74.7588   &  74.8408   &	  74.68418   \\
   12  &    2s$^2$2p$^5$3p	  &	$^3$D$  _{1}$	&   75.1061   &  75.1872   &	  75.04516   \\
   13  &    2s$^2$2p$^5$3p	  &	$^3$P$  _{1}$	&   75.4856   &  75.5648   &	  75.42378   \\
   14  &    2s$^2$2p$^5$3p	  &	$^1$D$  _{2}$	&   75.5343   &  75.6162   &	  75.46973   \\
   15  &    2s$^2$2p$^5$3p	  &	$^1$S$  _{0}$	&   76.6747   &  76.7670   &	  76.49453   \\
   16  &    2s$^2$2p$^5$3d	  &	$^3$P$^o_{0}$	&   77.3901   &  77.4471   &	  77.30283   \\
   17  &    2s$^2$2p$^5$3d	  &	$^3$P$^o_{1}$	&   77.4886   &  77.5468   &	  77.40108   \\
   18  &    2s$^2$2p$^5$3d	  &	$^3$P$^o_{2}$	&   77.6721   &  77.7263   &	  77.58429   \\
   19  &    2s$^2$2p$^5$3d	  &	$^3$F$^o_{4}$	&   77.6720   &  77.7326   &	  77.58456   \\
   20  &    2s$^2$2p$^5$3d	  &	$^3$F$^o_{3}$	&   77.7222   &  77.7778   &	  77.62885   \\
   21  &    2s$^2$2p$^5$3d	  &	$^1$D$^o_{2}$	&   77.8832   &  77.9411   &	  77.78970   \\
   22  &    2s$^2$2p$^5$3d	  &	$^3$D$^o_{3}$	&   77.9900   &  78.0527   &	  77.89623   \\
   23  &    2s$^2$2p$^5$3d	  &	$^3$D$^o_{1}$	&   78.4870   &  78.5407   &	  78.37643   \\
   24  &    2s$^2$2p$^5$3d	  &	$^3$F$^o_{2}$	&   79.2339   &  79.2971   &	  79.14201   \\
   25  &    2s$^2$2p$^5$3d	  &	$^3$D$^o_{2}$	&   79.3136   &  79.3733   &	  79.21323   \\
   26  &    2s$^2$2p$^5$3d	  &	$^1$F$^o_{3}$	&   79.3771   &  79.5361   &	  79.27460   \\
   27  &    2s$^2$2p$^5$3d	  &	$^1$P$^o_{1}$	&   79.9972   &  80.0433   &	  79.85892   \\
   28  &    2s2p$^6$3s  	  &	$^3$S$  _{1}$	&   82.6714   &  82.7901   &	  82.65656   \\
   29  &    2s2p$^6$3s  	  &	$^1$S$  _{0}$	&   83.2709   &  83.3917   &	  83.24921   \\
   30  &    2s2p$^6$3p  	  &	$^3$P$^o_{0}$	&   85.5424   &  85.6516   &	  85.54147   \\
   31  &    2s2p$^6$3p  	  &	$^3$P$^o_{1}$	&   85.5878   &  85.6984   &	  85.58796   \\
   32  &    2s2p$^6$3p  	  &	$^3$P$^o_{2}$	&   85.8917   &  86.0002   &	  85.89172   \\
   33  &    2s2p$^6$3p  	  &	$^1$P$^o_{1}$	&   86.0536   &  86.1689   &	  86.05657   \\
   34  &    2s2p$^6$3d  	  &	$^3$D$  _{1}$	&   89.5212   &  89.5967   &	  89.48917   \\
   35  &    2s2p$^6$3d  	  &	$^3$D$  _{2}$	&   89.5402   &  89.6158   &	  89.50820   \\
   36  &    2s2p$^6$3d  	  &	$^3$D$  _{3}$	&   89.5777   &  89.6530   &	  89.54538   \\
   37  &    2s2p$^6$3d  	  &	$^1$D$  _{2}$	&   90.0109   &  90.1029   &	  89.97868   \\
   38  &    2s$^2$2p$^5$4s	  &	$^3$P$^o_{2}$	&   95.5294   &  95.5800   &	  95.60201   \\
   39  &    2s$^2$2p$^5$4s	  &	$^1$P$^o_{1}$	&   95.5845   &  95.6373   &	  95.65949   \\
   40  &    2s$^2$2p$^5$4p	  &	$^3$S$  _{1}$	&   96.6490   &  96.6972   &	  96.71725   \\
   41  &    2s$^2$2p$^5$4p	  &	$^3$D$  _{2}$	&   96.7127   &  96.7619   &	  96.78351   \\
   42  &    2s$^2$2p$^5$4p	  &	$^3$D$  _{3}$	&   96.8157   &  96.8639   &	  96.88505   \\
   43  &    2s$^2$2p$^5$4p	  &	$^1$P$  _{1}$	&   96.8430   &  96.8919   &	  96.91355   \\
   44  &    2s$^2$2p$^5$4p	  &	$^3$P$  _{2}$	&   96.9026   &  96.9524   &	  96.97429   \\
   45  &    2s$^2$2p$^5$4s	  &	$^3$P$^o_{0}$	&   97.0570   &  97.1121   &	  97.13254   \\
\hline
\end{tabular}
\end{table}
\newpage
\begin{table}
\begin{tabular}{rllrrrrrrrrr} \hline
Index  &     Configuration        & Level               & GRASP        &   FAC1      &  FAC2 \\
\hline
   46  &    2s$^2$2p$^5$4s	  &	$^3$P$^o_{1}$	&   97.0837   &  97.1400   &	  97.16054   \\
   47  &    2s$^2$2p$^5$4p	  &	$^3$P$  _{0}$	&   97.2722   &  97.3121   &	  97.33091   \\
   48  &    2s$^2$2p$^5$4d	  &	$^3$P$^o_{0}$	&   98.1282   &  98.1747   &	  98.19264   \\
   49  &    2s$^2$2p$^5$4d	  &	$^3$P$^o_{1}$	&   98.1698   &  98.2154   &	  98.23391   \\
   50  &    2s$^2$2p$^5$4d	  &	$^3$F$^o_{4}$	&   98.2236   &  98.2687   &	  98.28776   \\
   51  &    2s$^2$2p$^5$4d	  &	$^3$F$^o_{3}$	&   98.2391   &  98.2691   &	  98.30323   \\
   52  &    2s$^2$2p$^5$4d	  &	$^3$P$^o_{2}$	&   98.2405   &  98.2832   &	  98.30276   \\
   53  &    2s$^2$2p$^5$4p	  &	$^3$D$  _{1}$	&   98.2169   &  98.2833   &	  98.28912   \\
   54  &    2s$^2$2p$^5$4d	  &	$^1$D$^o_{2}$	&   98.2988   &  98.3402   &	  98.36112   \\
   55  &    2s$^2$2p$^5$4d	  &	$^3$D$^o_{3}$	&   98.3389   &  98.3796   &	  98.40092   \\
   56  &    2s$^2$2p$^5$4p	  &	$^3$P$  _{1}$	&   98.3627   &  98.4162   &	  98.43573   \\
   57  &    2s$^2$2p$^5$4p	  &	$^1$D$  _{2}$	&   98.3833   &  98.4372   &	  98.45709   \\
   58  &    2s$^2$2p$^5$4d	  &	$^1$P$^o_{1}$	&   98.5765   &  98.6104   &	  98.63273   \\
   59  &    2s$^2$2p$^5$4p	  &	$^1$S$  _{0}$	&   98.5960   &  98.6370   &	  98.65317   \\
   60  &    2s$^2$2p$^5$4f	  &	$^3$D$  _{1}$	&   98.9161   &  98.9769   &	  98.99990   \\
   61  &    2s$^2$2p$^5$4f	  &	$^1$G$  _{4}$	&   98.9238   &    98.9864   &    99.00787   \\
   62  &    2s$^2$2p$^5$4f	  &	$^3$D$  _{2}$	&   98.9295   &    98.9910   &    99.01653   \\
   63  &    2s$^2$2p$^5$4f	  &	$^3$G$  _{5}$	&   98.9316   &    98.9943   &    99.01272   \\
   64  &    2s$^2$2p$^5$4f	  &	$^3$F$  _{3}$	&   98.9661   &    99.0296   &    99.05234   \\
   65  &    2s$^2$2p$^5$4f	  &	$^1$D$  _{2}$	&   98.9687   &    99.0363   &    99.05705   \\
   66  &    2s$^2$2p$^5$4f	  &	$^1$F$  _{3}$	&   98.9754   &    99.0399   &    99.06224   \\
   67  &    2s$^2$2p$^5$4f	  &	$^3$F$  _{4}$	&   98.9852   &    99.0500   &    99.07230   \\
   68  &    2s$^2$2p$^5$4d	  &	$^3$F$^o_{2}$	&   99.7552   &    99.8037   &    99.82186   \\
   69  &    2s$^2$2p$^5$4d	  &	$^3$D$^o_{2}$	&   99.7861   &    99.8341   &    99.85231   \\
   70  &    2s$^2$2p$^5$4d	  &	$^1$F$^o_{3}$	&   99.8147   &    99.8615   &    99.88031   \\
   71  &    2s$^2$2p$^5$4d	  &	$^3$D$^o_{1}$	&   99.9821   &   100.0200   &   100.04225   \\
   72  &    2s$^2$2p$^5$4f	  &	$^3$G$  _{3}$	&  100.4694   &   100.5370   &   100.55752   \\
   73  &    2s$^2$2p$^5$4f	  &	$^3$G$  _{4}$	&  100.4834   &   100.5510   &   100.57141   \\
   74  &    2s$^2$2p$^5$4f	  &	$^3$F$  _{2}$	&  100.4875   &   100.5580   &   100.57841   \\
   75  &    2s$^2$2p$^5$4f	  &	$^3$D$  _{3}$	&  100.4906   &   100.5580   &   100.57952   \\
   76  &    2s$^2$2p$^5$5s	  &	$^3$P$^o_{2}$	&  106.2758   &   106.3310   &   106.35044   \\
   77  &    2s$^2$2p$^5$5s	  &	$^1$P$^o_{1}$	&  106.3043   &   106.3590   &   106.37932   \\
   78  &    2s$^2$2p$^5$5p	  &	$^3$S$  _{1}$	&  106.8108   &   106.8650   &   106.88153   \\
   79  &    2s$^2$2p$^5$5p	  &	$^3$D$  _{2}$	&  106.8680   &   106.9210   &   106.94092   \\
   80  &    2s$^2$2p$^5$5p	  &	$^3$D$  _{3}$	&  106.9194   &   106.9730   &   106.99187   \\
   81  &    2s$^2$2p$^5$5p	  &	$^1$P$  _{1}$	&  106.9290   &   106.9820   &   107.00132   \\
   82  &    2s$^2$2p$^5$5p	  &	$^3$P$  _{2}$	&  106.9625   &   107.0150   &   107.03518   \\
   83  &    2s$^2$2p$^5$5p	  &	$^1$S$  _{0}$	&  107.1485   &   107.1940   &   107.21065   \\
   84  &    2s2p$^6$4s  	  &	$^3$S$  _{1}$	&  107.4520   &   107.4750   &   107.49048   \\
   85  &    2s$^2$2p$^5$5d	  &	$^3$P$^o_{0}$	&  107.5565   &   107.6170   &   107.63197   \\
   86  &    2s$^2$2p$^5$5d	  &	$^3$P$^o_{1}$	&  107.5785   &   107.6370   &   107.65339   \\
   87  &    2s$^2$2p$^5$5d	  &	$^3$F$^o_{4}$	&  107.6066   &   107.6480   &   107.68114   \\
   88  &    2s$^2$2p$^5$5d	  &	$^3$F$^o_{3}$	&  107.6141   &   107.6640   &   107.68710   \\
   89  &    2s$^2$2p$^5$5d	  &	$^3$P$^o_{2}$	&  107.6138   &   107.6690   &   107.68814   \\
   90  &    2s$^2$2p$^5$5d	  &	$^1$D$^o_{2}$	&  107.6414   &   107.6710   &   107.71408   \\
\hline
\end{tabular}
\end{table}
\newpage
\begin{table}
\begin{tabular}{rllrrrrrrrrr} \hline
Index  &     Configuration        & Level               & GRASP        &   FAC1      &  FAC2 \\
\hline
   91  &    2s2p$^6$4s  	  &	$^1$S$  _{0}$	&  107.6356   &   107.6950   &   107.66429   \\
   92  &    2s$^2$2p$^5$5d	  &	$^3$D$^o_{3}$	&  107.6611   &   107.7140   &   107.73308   \\
   93  &    2s$^2$2p$^5$5d	  &	$^1$P$^o_{1}$	&  107.7641   &   107.8130   &   107.83283   \\
   94  &    2s$^2$2p$^5$5s	  &	$^3$P$^o_{0}$	&  107.7971   &   107.8560   &   107.87410   \\
   95  &    2s$^2$2p$^5$5s	  &	$^3$P$^o_{1}$	&  107.8348   &   107.8900   &   107.90797   \\
   96  &    2s$^2$2p$^5$5f	  &	$^3$D$  _{1}$	&  107.9400   &   108.0060   &   108.02713   \\
   97  &    2s$^2$2p$^5$5f	  &	$^3$D$  _{2}$	&  107.9508   &   108.0180   &   108.03856   \\
   98  &    2s$^2$2p$^5$5f	  &	$^3$G$  _{4}$	&  107.9545   &   108.0220   &   108.04248   \\
   99  &    2s$^2$2p$^5$5f	  &	$^3$G$  _{5}$	&  107.9565   &   108.0240   &   108.04405   \\
  100  &    2s$^2$2p$^5$5f	  &	$^3$D$  _{3}$	&  107.9710   &   108.0390   &   108.05925   \\
  101  &    2s$^2$2p$^5$5f	  &	$^1$D$  _{2}$	&  107.9778   &   108.0400   &   108.06748   \\
  102  &    2s$^2$2p$^5$5f	  &	$^1$F$  _{3}$	&  107.9792   &   108.0440   &   108.06792   \\
  103  &    2s$^2$2p$^5$5f	  &	$^3$F$  _{4}$	&  107.9848   &   108.0470   &   108.07363   \\
  104  &    2s$^2$2p$^5$5p	  &	$^3$D$  _{3}$	&  108.3864   &   108.0480   &   108.46130   \\
  105  &    2s$^2$2p$^5$5p	  &	$^3$P$  _{1}$	&  108.4674   &   108.0530   &   108.54401   \\
  106  &    2s$^2$2p$^5$5p	  &	$^1$D$  _{2}$	&  108.4687   &   108.0550   &   108.54325   \\
  107  &    2s$^2$2p$^5$5p	  &	$^3$P$  _{0}$	&  108.5485   &   108.0580   &   108.61346   \\
  108  &    2s2p$^6$4p  	  &	$^3$P$^o_{0}$	&  108.6111   &   108.0610   &   108.64568   \\
  109  &    2s2p$^6$4p  	  &	$^3$P$^o_{1}$	&  108.6158   &   108.0630   &   108.65070   \\
  110  &    2s2p$^6$4p  	  &	$^3$P$^o_{2}$	&  108.7342   &   108.0690   &   108.76862   \\
  111  &    2s2p$^6$4p  	  &	$^1$P$^o_{1}$	&  108.7914   &   108.0720   &   108.82734   \\
  112  &    2s$^2$2p$^5$5d	  &	$^3$F$^o_{2}$	&  109.1362   &   108.4430   &   109.21307   \\
  113  &    2s$^2$2p$^5$5d	  &	$^3$D$^o_{2}$	&  109.1553   &   108.5250   &   109.23176   \\
  114  &    2s$^2$2p$^5$5d	  &	$^1$F$^o_{3}$	&  109.1658   &   108.5260   &   109.24195   \\
  115  &    2s$^2$2p$^5$5d	  &	$^3$D$^o_{1}$	&  109.2410   &   108.6000   &   109.31123   \\
  116  &    2s$^2$2p$^5$5f	  &	$^3$G$  _{3}$	&  109.4909   &   108.6300   &   109.58144   \\
  117  &    2s$^2$2p$^5$5f	  &	$^3$F$  _{3}$	&  109.4927   &   108.6350   &   109.58350   \\
  118  &    2s$^2$2p$^5$5f	  &	$^3$F$  _{2}$	&  109.4972   &   108.7540   &   109.58903   \\
  119  &    2s$^2$2p$^5$5f	  &	$^3$G$  _{4}$	&  109.4997   &   108.8100   &   109.59068   \\
  120  &    2s2p$^6$4d  	  &	$^3$D$  _{1}$	&  110.1045   &   109.1970   &   110.12934   \\
  121  &    2s2p$^6$4d  	  &	$^3$D$  _{2}$	&  110.1152   &   109.2160   &   110.14050 \\
  122  &    2s2p$^6$4d		  &     $^3$D$  _{3}$   &  110.1349   &   109.2250   &   110.16087 \\
  123  &    2s2p$^6$4d		  &     $^1$D$  _{2}$   &  110.2809   &   109.2940   &   110.30125 \\
  124  &    2s2p$^6$4f		  &     $^3$F$^o_{2}$   &  110.7921   &   109.5630   &   110.84024 \\
  125  &    2s2p$^6$4f		  &     $^3$F$^o_{3}$   &  110.7941   &   109.5650   &   110.84754 \\
  126  &    2s2p$^6$4f		  &     $^3$F$^o_{4}$   &  110.8017   &   109.5710   &   110.85810 \\
  127  &    2s2p$^6$4f		  &     $^1$F$^o_{3}$   &  110.8157   &   109.5720   &   110.86835 \\
\hline				  				        			
\end{tabular}	
									            			    
\begin {flushleft}									            			    
\begin{tabbing} 									      	
aaaaaaaaaaaaaaaaaaaaaaaaaaaaaaaaaaaa\= \kill						      	       
GRASP: Earlier results of Singh and Aggarwal \cite{sa}  with the {\sc grasp} code  \\			
FAC1: Earlier results of Singh and Aggarwal \cite{sa}  with the {\sc fac} code  \\ 			
FAC2: Present results  with the {\sc fac} code  			      		      	
\end{tabbing}										      	
\end {flushleft}						
\end{table}

\begin{table}
\caption{Comparison of energies (in Ryd) for 127 levels of  Zn XXI.} 
\begin{tabular}{rllrrrrr} \hline
Index  &     Configuration        & Level               & GRASP        &    FAC1    &  FAC2       \\
 \hline
    1  &    2s$^2$2p$^6$	  &	$^1$S$  _{0}$	&    0.0000   &   0.0000    &	   0.00000 \\
    2  &    2s$^2$2p$^5$3s	  &	$^3$P$^o_{2}$	&   77.0829   &  77.1622    &	  77.02974 \\
    3  &    2s$^2$2p$^5$3s	  &	$^1$P$^o_{1}$	&   77.2657   &  77.3546    &	  77.20457 \\
    4  &    2s$^2$2p$^5$3s	  &	$^3$P$^o_{0}$	&   78.8614   &  78.9435    &	  78.80181 \\
    5  &    2s$^2$2p$^5$3s	  &	$^3$P$^o_{1}$	&   78.9649   &  79.0539    &	  78.90022 \\
    6  &    2s$^2$2p$^5$3p	  &	$^3$S$  _{1}$	&   79.9196   &  79.9897    &	  79.87126 \\
    7  &    2s$^2$2p$^5$3p	  &	$^3$D$  _{2}$	&   80.1804   &  80.2601    &	  80.12322 \\
    8  &    2s$^2$2p$^5$3p	  &	$^3$D$  _{3}$	&   80.4828   &  80.5583    &	  80.42992 \\
    9  &    2s$^2$2p$^5$3p	  &	$^1$P$  _{1}$	&   80.5515   &  80.6322    &	  80.49657 \\
   10  &    2s$^2$2p$^5$3p	  &	$^3$P$  _{2}$	&   80.7564   &  80.8381    &	  80.69710 \\
   11  &    2s$^2$2p$^5$3p	  &	$^3$P$  _{0}$	&   81.4461   &  81.5289    &	  81.37004 \\
   12  &    2s$^2$2p$^5$3p	  &	$^3$D$  _{1}$	&   81.8956   &  81.9773    &	  81.83533 \\
   13  &    2s$^2$2p$^5$3p	  &	$^3$P$  _{1}$	&   82.3366   &  82.4164    &	  82.27576 \\
   14  &    2s$^2$2p$^5$3p	  &	$^1$D$  _{2}$	&   82.3920   &  82.4746    &	  82.32825 \\
   15  &    2s$^2$2p$^5$3p	  &	$^1$S$  _{0}$	&   83.4967   &  83.5888    &	  83.32059 \\
   16  &    2s$^2$2p$^5$3d	  &	$^3$P$^o_{0}$	&   84.1579   &  84.2146    &	  84.07169 \\
   17  &    2s$^2$2p$^5$3d	  &	$^3$P$^o_{1}$	&   84.2661   &  84.3240    &	  84.17962 \\
   18  &    2s$^2$2p$^5$3d	  &	$^3$P$^o_{2}$	&   84.4671   &  84.5172    &	  84.38020 \\
   19  &    2s$^2$2p$^5$3d	  &	$^3$F$^o_{4}$	&   84.4631   &  84.5274    &	  84.37681 \\
   20  &    2s$^2$2p$^5$3d	  &	$^3$F$^o_{3}$	&   84.5079   &  84.5635    &	  84.41571 \\
   21  &    2s$^2$2p$^5$3d	  &	$^1$D$^o_{2}$	&   84.6811   &  84.7391    &	  84.58873 \\
   22  &    2s$^2$2p$^5$3d	  &	$^3$D$^o_{3}$	&   84.8000   &  84.8624    &	  84.70740 \\
   23  &    2s$^2$2p$^5$3d	  &	$^3$D$^o_{1}$	&   85.3457   &  85.3986    &	  85.23471 \\
   24  &    2s$^2$2p$^5$3d	  &	$^3$F$^o_{2}$	&   86.2698   &  86.3332    &	  86.17882 \\
   25  &    2s$^2$2p$^5$3d	  &	$^3$D$^o_{2}$	&   86.3610   &  86.4209    &	  86.26147 \\
   26  &    2s$^2$2p$^5$3d	  &	$^1$F$^o_{3}$	&   86.4317   &  86.4909    &	  86.32999 \\
   27  &    2s$^2$2p$^5$3d	  &	$^1$P$^o_{1}$	&   87.0583   &  87.1054    &	  86.92262 \\
   28  &    2s2p$^6$3s  	  &	$^3$S$  _{1}$	&   89.7723   &  89.8922    &	  89.75857 \\
   29  &    2s2p$^6$3s  	  &	$^1$S$  _{0}$	&   90.4013   &  90.5230    &	  90.38050 \\
   30  &    2s2p$^6$3p  	  &	$^3$P$^o_{0}$	&   92.7900   &  92.9003    &	  92.78983 \\
   31  &    2s2p$^6$3p  	  &	$^3$P$^o_{1}$	&   92.8395   &  92.9512    &	  92.84050 \\
   32  &    2s2p$^6$3p  	  &	$^3$P$^o_{2}$	&   93.2025   &  93.3119    &	  93.20325 \\
   33  &    2s2p$^6$3p  	  &	$^1$P$^o_{1}$	&   93.3696   &  93.4859    &	  93.37355 \\
   34  &    2s2p$^6$3d  	  &	$^3$D$  _{1}$	&   97.0103   &  97.0861    &	  96.97910 \\
   35  &    2s2p$^6$3d  	  &	$^3$D$  _{2}$	&   97.0332   &  97.1092    &	  97.00204 \\
   36  &    2s2p$^6$3d  	  &	$^3$D$  _{3}$	&   97.0790   &  97.1547    &	  97.04758 \\
   37  &    2s2p$^6$3d  	  &	$^1$D$  _{2}$	&   97.5360   &  97.6286    &	  97.50471 \\
   38  &    2s$^2$2p$^5$4s	  &	$^3$P$^o_{2}$	&  104.2275   & 104.2790    &	 104.30032 \\
   39  &    2s$^2$2p$^5$4s	  &	$^1$P$^o_{1}$	&  104.2857   & 104.3390    &	 104.36092 \\
   40  &    2s$^2$2p$^5$4p	  &	$^3$S$  _{1}$	&  105.4109   & 105.4600    &	 105.47927 \\
   41  &    2s$^2$2p$^5$4p	  &	$^3$D$  _{2}$	&  105.4742   & 105.5240    &	 105.54500 \\
   42  &    2s$^2$2p$^5$4p	  &	$^3$D$  _{3}$	&  105.6007   & 105.6500    &	 105.67009 \\
   43  &    2s$^2$2p$^5$4p	  &	$^1$P$  _{1}$	&  105.6269   & 105.6760    &	 105.69740 \\
   44  &    2s$^2$2p$^5$4p	  &	$^3$P$  _{2}$	&  105.6926   & 105.7430    &	 105.76432 \\
   45  &    2s$^2$2p$^5$4s	  &	$^3$P$^o_{0}$	&  106.0096   & 106.0660    &	 106.08551 \\
\hline
\end{tabular}
\end{table}
\newpage
\begin{table}
\begin{tabular}{rllrrrrrrrrr} \hline
Index  &     Configuration        & Level               & GRASP        &   FAC1      &  FAC2 \\
\hline
   46  &    2s$^2$2p$^5$4p	  &	$^3$P$  _{0}$	&  106.0372   & 106.0950    &	 106.14487 \\
   47  &    2s$^2$2p$^5$4s	  &	$^3$P$^o_{1}$	&  106.0873   & 106.1270    &	 106.11443 \\
   48  &    2s$^2$2p$^5$4d	  &	$^3$P$^o_{0}$	&  106.9794   & 107.0270    &	 107.04406 \\
   49  &    2s$^2$2p$^5$4d	  &	$^3$P$^o_{1}$	&  107.0244   & 107.0710    &	 107.08862 \\
   50  &    2s$^2$2p$^5$4d	  &	$^3$F$^o_{4}$	&  107.0839   & 107.1300    &	 107.14819 \\
   51  &    2s$^2$2p$^5$4d	  &	$^3$F$^o_{3}$	&  107.0977   & 107.1410    &	 107.16050 \\
   52  &    2s$^2$2p$^5$4d	  &	$^3$P$^o_{2}$	&  107.0998   & 107.1450    &	 107.16348 \\
   53  &    2s$^2$2p$^5$4d	  &	$^1$D$^o_{2}$	&  107.1607   & 107.2030    &	 107.22317 \\
   54  &    2s$^2$2p$^5$4d	  &	$^3$D$^o_{3}$	&  107.2059   & 107.2470    &	 107.26792 \\
   55  &    2s$^2$2p$^5$4p	  &	$^3$D$  _{1}$	&  107.2311   & 107.2840    &	 107.30363 \\
   56  &    2s$^2$2p$^5$4d	  &	$^1$P$^o_{1}$	&  107.4028   & 107.4570    &	 107.47601 \\
   57  &    2s$^2$2p$^5$4p	  &	$^3$P$  _{1}$	&  107.4247   & 107.4800    &	 107.51529 \\
   58  &    2s$^2$2p$^5$4p	  &	$^1$D$  _{2}$	&  107.4593   & 107.4940    &	 107.49883 \\
   59  &    2s$^2$2p$^5$4p	  &	$^1$S$  _{0}$	&  107.6192   & 107.6620    &	 107.67708 \\
   60  &    2s$^2$2p$^5$4f	  &	$^3$D$  _{1}$	&  107.8197   & 107.8810    &	 107.90350 \\
   61  &    2s$^2$2p$^5$4f	  &	$^1$G$  _{4}$	&   107.8297   & 107.8930    &    107.91365 \\
   62  &    2s$^2$2p$^5$4f	  &	$^3$D$  _{2}$	&   107.8378   & 107.8990    &    107.92260 \\
   63  &    2s$^2$2p$^5$4f	  &	$^3$G$  _{5}$	&   107.8372   & 107.9010    &    107.92036 \\
   64  &    2s$^2$2p$^5$4f	  &	$^3$F$  _{3}$	&   107.8752   & 107.9390    &    107.96143 \\
   65  &    2s$^2$2p$^5$4f	  &	$^1$D$  _{2}$	&   107.8781   & 107.9470    &    107.96655 \\
   66  &    2s$^2$2p$^5$4f	  &	$^1$F$  _{3}$	&   107.8853   & 107.9510    &    107.97209 \\
   67  &    2s$^2$2p$^5$4f	  &	$^3$F$  _{4}$	&   107.8969   & 107.9620    &    107.98395 \\
   68  &    2s$^2$2p$^5$4d	  &	$^3$F$^o_{2}$	&   108.8657   & 108.9160    &    108.93279 \\
   69  &    2s$^2$2p$^5$4d	  &	$^3$D$^o_{2}$	&   108.9023   & 108.9520    &    108.96901 \\
   70  &    2s$^2$2p$^5$4d	  &	$^1$F$^o_{3}$	&   108.9329   & 108.9810    &    108.99892 \\
   71  &    2s$^2$2p$^5$4d	  &	$^3$D$^o_{1}$	&   109.1018   & 109.1430    &    109.16225 \\
   72  &    2s$^2$2p$^5$4f	  &	$^3$G$  _{3}$	&   109.6309   & 109.7000    &    109.71927 \\
   73  &    2s$^2$2p$^5$4f	  &	$^3$G$  _{4}$	&   109.6476   & 109.7170    &    109.73579 \\
   74  &    2s$^2$2p$^5$4f	  &	$^3$F$  _{2}$	&   109.6499   & 109.7220    &    109.74110 \\
   75  &    2s$^2$2p$^5$4f	  &	$^3$D$  _{3}$	&   109.6543   & 109.7230    &    109.74351 \\
   76  &    2s$^2$2p$^5$5s	  &	$^3$P$^o_{2}$	&   116.0031   & 116.0590    &    116.07773 \\
   77  &    2s$^2$2p$^5$5s	  &	$^1$P$^o_{1}$	&   116.0333   & 116.0900    &    116.10839 \\
   78  &    2s$^2$2p$^5$5p	  &	$^3$S$  _{1}$	&   116.5293   & 116.5800    &    116.59354 \\
   79  &    2s$^2$2p$^5$5p	  &	$^3$D$  _{2}$	&   116.6284   & 116.6820    &    116.70123 \\
   80  &    2s$^2$2p$^5$5p	  &	$^1$P$  _{1}$	&   116.6920   & 116.7450    &    116.76329 \\
   81  &    2s$^2$2p$^5$5p	  &	$^3$D$  _{3}$	&   116.6915   & 116.7460    &    116.76395 \\
   82  &    2s$^2$2p$^5$5p	  &	$^3$P$  _{2}$	&   116.7372   & 116.7910    &    116.80986 \\
   83  &    2s$^2$2p$^5$5p	  &	$^1$S$  _{0}$	&   116.9031   & 116.9370    &    116.96329 \\
   84  &    2s2p$^6$4s  	  &	$^3$S$  _{1}$	&   116.9081   & 116.9460    &    116.95404 \\
   85  &    2s2p$^6$4s  	  &	$^1$S$  _{0}$	&   117.0788   & 117.0960    &    117.10867 \\
   86  &    2s$^2$2p$^5$5d	  &	$^3$P$^o_{0}$	&   117.3596   & 117.4210    &    117.43507 \\
   87  &    2s$^2$2p$^5$5d	  &	$^3$P$^o_{1}$	&   117.3839   & 117.4440    &    117.45881 \\
   88  &    2s$^2$2p$^5$5d	  &	$^3$F$^o_{4}$	&   117.4162   & 117.4750    &    117.49099 \\
   89  &    2s$^2$2p$^5$5d	  &	$^3$F$^o_{3}$	&   117.4217   & 117.4780    &    117.49515 \\
   90  &    2s$^2$2p$^5$5d	  &	$^3$P$^o_{2}$	&   117.4232   & 117.4810    &    117.49727 \\
\hline
\end{tabular}
\end{table}
\newpage
\begin{table}
\begin{tabular}{rllrrrrrrrrr} \hline
Index  &     Configuration        & Level               & GRASP        &   FAC1      &  FAC2 \\
\hline
   91  &    2s$^2$2p$^5$5d	  &	$^1$D$^o_{2}$	&   117.4517   & 117.5070    &    117.52450 \\
   92  &    2s$^2$2p$^5$5d	  &	$^3$D$^o_{3}$	&   117.4739   & 117.5280    &    117.54601 \\
   93  &    2s$^2$2p$^5$5d	  &	$^1$P$^o_{1}$	&   117.5966   & 117.6430    &    117.66148 \\
   94  &    2s$^2$2p$^5$5f	  &	$^3$D$  _{1}$	&   117.7695   & 117.8100    &    117.85645 \\
   95  &    2s$^2$2p$^5$5f	  &	$^3$D$  _{2}$	&   117.7819   & 117.8370    &    117.86942 \\
   96  &    2s$^2$2p$^5$5f	  &	$^3$G$  _{4}$	&   117.7517   & 117.8470    &    117.87419 \\
   97  &    2s$^2$2p$^5$5f	  &	$^3$G$  _{5}$	&   117.7864   & 117.8500    &    117.87666 \\
   98  &    2s$^2$2p$^5$5s	  &	$^3$P$^o_{0}$	&   117.7893   & 117.8550    &    117.82523 \\
   99  &    2s$^2$2p$^5$5f	  &	$^3$D$  _{3}$	&   117.8040   & 117.8580    &    117.89209 \\
  100  &    2s$^2$2p$^5$5f	  &	$^1$D$  _{2}$	&   117.8117   & 117.8720    &    117.90127 \\
  101  &    2s$^2$2p$^5$5f	  &	$^1$F$  _{3}$	&   117.8129   & 117.8760    &    117.90144 \\
  102  &    2s$^2$2p$^5$5f	  &	$^3$F$  _{4}$	&   117.8195   & 117.8800    &    117.90820 \\
  103  &    2s$^2$2p$^5$5s	  &	$^3$P$^o_{1}$	&   117.7886   & 117.8820    &    117.86364 \\
  104  &    2s2p$^6$4p  	  &	$^3$P$^o_{1}$	&   118.0939   & 117.8830    &    118.13043 \\
  105  &    2s2p$^6$4p  	  &	$^3$P$^o_{0}$	&   118.1060   & 117.8890    &    118.14479 \\
  106  &    2s2p$^6$4p  	  &	$^3$P$^o_{2}$	&   118.2282   & 117.8920    &    118.26240 \\
  107  &    2s2p$^6$4p  	  &	$^1$P$^o_{1}$	&   118.2918   & 117.8960    &    118.32829 \\
  108  &    2s$^2$2p$^5$5p	  &	$^3$D$  _{3}$	&   118.3992   & 117.8980    &    118.47430 \\
  109  &    2s$^2$2p$^5$5p	  &	$^3$P$  _{1}$	&   118.4932   & 117.9010    &    118.56895 \\
  110  &    2s$^2$2p$^5$5p	  &	$^1$D$  _{2}$	&   118.4940   & 117.9080    &    118.57005 \\
  111  &    2s$^2$2p$^5$5p	  &	$^3$P$  _{0}$	&   118.5667   & 117.9110    &    118.63262 \\
  112  &    2s$^2$2p$^5$5d	  &	$^3$F$^o_{2}$	&   119.1983   & 118.1150    &    119.27555 \\
  113  &    2s$^2$2p$^5$5d	  &	$^3$D$^o_{2}$	&   119.2183   & 118.1290    &    119.29566 \\
  114  &    2s$^2$2p$^5$5d	  &	$^1$F$^o_{3}$	&   119.2316   & 118.2480    &    119.30828 \\
  115  &    2s$^2$2p$^5$5d	  &	$^3$D$^o_{1}$	&   119.3062   & 118.3120    &    119.37708 \\
  116  &    2s$^2$2p$^5$5f	  &	$^3$F$  _{3}$	&   119.5431   & 118.4580    &    119.61725 \\
  117  &    2s$^2$2p$^5$5f	  &	$^3$F$  _{2}$	&   119.5597   & 118.5520    &    119.63652 \\
  118  &    2s$^2$2p$^5$5f	  &	$^3$G$  _{3}$	&   119.5790   & 118.5530    &    119.67037 \\
  119  &    2s$^2$2p$^5$5f	  &	$^3$G$  _{4}$	&   119.5881   & 118.6200    &    119.67924 \\
  120  &    2s2p$^6$4d  	  &	$^3$D$  _{1}$	&   119.6672   & 119.2610    &    119.69228 \\
  121  &    2s2p$^6$4d  	  &	$^3$D$  _{2}$	&   119.7023  & 119.2810    &	 119.74253 \\
  122  &    2s2p$^6$4d		  &     $^3$D$  _{3}$   &   119.7378  & 119.2930    &	 119.78004 \\
  123  &    2s2p$^6$4d		  &     $^1$D$  _{2}$   &   119.8555  & 119.3620    &	 119.87640 \\
  124  &    2s2p$^6$4f		  &     $^3$F$^o_{2}$   &   120.3999  & 119.6030    &	 120.45145 \\
  125  &    2s2p$^6$4f		  &     $^3$F$^o_{3}$   &   120.4023  & 119.6220    &	 120.46340 \\
  126  &    2s2p$^6$4f		  &     $^3$F$^o_{4}$   &   120.4117  & 119.6530    &	 120.47974 \\
  127  &    2s2p$^6$4f		  &     $^1$F$^o_{3}$   &   120.4269  & 119.6620    &	 120.48666 \\
\hline				  				        				   
\end{tabular}	
									            			    
\begin {flushleft}									            			    
\begin{tabbing} 									      	
aaaaaaaaaaaaaaaaaaaaaaaaaaaaaaaaaaaa\= \kill						      	       
GRASP: Earlier results of Singh and Aggarwal \cite{sa} with the {\sc grasp} code  \\			
FAC1: Earlier results of Singh and Aggarwal \cite{sa} with the {\sc fac} code  \\ 			
FAC2: Present results  with the {\sc fac} code  			      		      	
\end{tabbing}										      	
\end {flushleft}								
\end{table}

\newpage
\begin{table}
\caption{Comparison of energies (in Ryd) for 127 levels of  Ga XXII.} 
\begin{tabular}{rllrrrrr} \hline
Index  &     Configuration        & Level               & GRASP        &    FAC1    &  FAC2       \\
 \hline
    1  &    2s$^2$2p$^6$	  &	$^1$S$  _{0}$	&     0.0000   &    0.0000   &     0.00000  \\
    2  &    2s$^2$2p$^5$3s	  &	$^3$P$^o_{2}$	&    83.7426   &   83.8227   &    83.69079  \\
    3  &    2s$^2$2p$^5$3s	  &	$^1$P$^o_{1}$	&    83.9348   &   84.0246   &    83.87482  \\
    4  &    2s$^2$2p$^5$3s	  &	$^3$P$^o_{0}$	&    85.8058   &   85.8890   &    85.74715  \\
    5  &    2s$^2$2p$^5$3s	  &	$^3$P$^o_{1}$	&    85.9122   &   86.0023   &    85.84867  \\
    6  &    2s$^2$2p$^5$3p	  &	$^3$S$  _{1}$	&    86.7349   &   86.8059   &    86.68729  \\
    7  &    2s$^2$2p$^5$3p	  &	$^3$D$  _{2}$	&    86.9911   &   87.0714   &    86.93499  \\
    8  &    2s$^2$2p$^5$3p	  &	$^3$D$  _{3}$	&    87.3592   &   87.4354   &    87.30757  \\
    9  &    2s$^2$2p$^5$3p	  &	$^1$P$  _{1}$	&    87.4221   &   87.5033   &    87.36857  \\
   10  &    2s$^2$2p$^5$3p	  &	$^3$P$  _{2}$	&    87.6461   &   87.7284   &    87.58796  \\ 
   11  &    2s$^2$2p$^5$3p	  &	$^3$P$  _{0}$	&    88.4146   &   88.4982   &    88.33686  \\
   12  &    2s$^2$2p$^5$3p	  &	$^3$D$  _{1}$	&    88.9853   &   89.0679   &    88.92588  \\
   13  &    2s$^2$2p$^5$3p	  &	$^3$P$  _{1}$	&    89.4964   &   89.5769   &    89.43643  \\
   14  &    2s$^2$2p$^5$3p	  &	$^1$D$  _{2}$	&    89.5583   &   89.6419   &    89.49547  \\
   15  &    2s$^2$2p$^5$3p	  &	$^1$S$  _{0}$	&    90.6161   &   90.7082   &    90.44428  \\
   16  &    2s$^2$2p$^5$3d	  &	$^3$P$^o_{0}$	&    91.2036   &   91.2602   &    91.11856  \\
   17  &    2s$^2$2p$^5$3d	  &	$^3$P$^o_{1}$	&    91.3219   &   91.3798   &    91.23651  \\
   18  &    2s$^2$2p$^5$3d	  &	$^3$P$^o_{2}$	&    91.5411   &   91.5879   &    91.45519  \\
   19  &    2s$^2$2p$^5$3d	  &	$^3$F$^o_{4}$	&    91.5338   &   91.6013   &    91.44879  \\
   20  &    2s$^2$2p$^5$3d	  &	$^3$F$^o_{3}$	&    91.5713   &   91.6270   &    91.48034  \\
   21  &    2s$^2$2p$^5$3d	  &	$^1$D$^o_{2}$	&    91.7574   &   91.8157   &    91.66630  \\
   22  &    2s$^2$2p$^5$3d	  &	$^3$D$^o_{3}$	&    91.8898   &   91.9525   &    91.79834  \\
   23  &    2s$^2$2p$^5$3d	  &	$^3$D$^o_{1}$	&    92.4846   &   92.5370   &    92.37331  \\
   24  &    2s$^2$2p$^5$3d	  &	$^3$F$^o_{2}$	&    93.6143   &   93.6780   &    93.52431  \\
   25  &    2s$^2$2p$^5$3d	  &	$^3$D$^o_{2}$	&    93.7190   &   93.7793   &    93.62042  \\
   26  &    2s$^2$2p$^5$3d	  &	$^1$F$^o_{3}$	&    93.7968   &   93.8564   &    93.69603  \\
   27  &    2s$^2$2p$^5$3d	  &	$^1$P$^o_{1}$	&    94.4270   &   94.4753   &    94.29389  \\
   28  &    2s2p$^6$3s  	  &	$^3$S$  _{1}$	&    97.1756   &   97.2969   &    97.16302  \\
   29  &    2s2p$^6$3s  	  &	$^1$S$  _{0}$	&    97.8344   &   97.9571   &    97.81450  \\
   30  &    2s2p$^6$3p  	  &	$^3$P$^o_{0}$	&   100.3409   &  100.4520   &   100.34146  \\
   31  &    2s2p$^6$3p  	  &	$^3$P$^o_{1}$	&   100.3943   &  100.5070   &   100.39630  \\
   32  &    2s2p$^6$3p  	  &	$^3$P$^o_{2}$	&   100.8246   &  100.9350   &   100.82625  \\
   33  &    2s2p$^6$3p  	  &	$^1$P$^o_{1}$	&   100.9970   &  101.1140   &   101.00202  \\
   34  &    2s2p$^6$3d  	  &	$^3$D$  _{1}$	&   104.8107   &  104.8870   &   104.78048  \\
   35  &    2s2p$^6$3d  	  &	$^3$D$  _{2}$	&   104.8380   &  104.9150   &   104.80786  \\
   36  &    2s2p$^6$3d  	  &	$^3$D$  _{3}$	&   104.8936   &  104.9700   &   104.86314  \\
   37  &    2s2p$^6$3d  	  &	$^1$D$  _{2}$	&   105.3738   &  105.4670   &   105.34357  \\
   38  &    2s$^2$2p$^5$4s	  &	$^3$P$^o_{2}$	&   113.2984   &  113.3510   &   113.37141  \\
   39  &    2s$^2$2p$^5$4s	  &	$^1$P$^o_{1}$	&   113.3597   &  113.4140   &   113.43512  \\
   40  &    2s$^2$2p$^5$4p	  &	$^3$S$  _{1}$	&   114.5460   &  114.5960   &   114.61448  \\
   41  &    2s$^2$2p$^5$4p	  &	$^3$D$  _{2}$	&   114.6088   &  114.6590   &   114.67960  \\
   42  &    2s$^2$2p$^5$4p	  &	$^3$D$  _{3}$	&   114.7622   &  114.8120   &   114.83161  \\
   43  &    2s$^2$2p$^5$4p	  &	$^1$P$  _{1}$	&   114.7872   &  114.8380   &   114.85764  \\
   44  &    2s$^2$2p$^5$4p	  &	$^3$P$  _{2}$	&   114.8591   &  114.9100   &   114.93084  \\
   45  &    2s$^2$2p$^5$4p	  &	$^3$P$  _{0}$	&   115.2783   &  115.3180   &   115.33489  \\
\hline
\end{tabular}
\end{table}
\newpage
\begin{table}
\begin{tabular}{rllrrrrrrrrr} \hline
Index  &     Configuration        & Level               & GRASP        &   FAC1      &  FAC2 \\
\hline
   46  &    2s$^2$2p$^5$4s	  &	$^3$P$^o_{0}$	&   115.3657   &  115.4230   &   115.44212  \\
   47  &    2s$^2$2p$^5$4s	  &	$^3$P$^o_{1}$	&   115.3941   &  115.4530   &   115.47182  \\
   48  &    2s$^2$2p$^5$4d	  &	$^3$P$^o_{0}$	&   116.2073   &  116.2560   &   116.27206  \\
   49  &    2s$^2$2p$^5$4d	  &	$^3$P$^o_{1}$	&   116.2556   &  116.3030   &   116.32000  \\
   50  &    2s$^2$2p$^5$4d	  &	$^3$F$^o_{4}$	&   116.3215   &  116.3680   &   116.38595  \\
   51  &    2s$^2$2p$^5$4d	  &	$^3$F$^o_{3}$	&   116.3313   &  116.3760   &   116.39421  \\
   52  &    2s$^2$2p$^5$4d	  &	$^3$P$^o_{2}$	&   116.3374   &  116.3830   &   116.40111  \\
   53  &    2s$^2$2p$^5$4d	  &	$^1$D$^o_{2}$	&   116.3996   &  116.4430   &   116.46217  \\
   54  &    2s$^2$2p$^5$4d	  &	$^3$D$^o_{3}$	&   116.4501   &  116.4920   &   116.51226  \\
   55  &    2s$^2$2p$^5$4p	  &	$^3$D$  _{1}$	&   116.6492   &  116.7040   &   116.72211  \\
   56  &    2s$^2$2p$^5$4d	  &	$^1$P$^o_{1}$	&   116.7194   &  116.7540   &   116.77510  \\
   57  &    2s$^2$2p$^5$4p	  &	$^3$P$  _{1}$	&   116.8501   &  116.9060   &   116.92368  \\
   58  &    2s$^2$2p$^5$4p	  &	$^1$D$  _{2}$	&   116.8734   &  116.9300   &   116.94786  \\
   59  &    2s$^2$2p$^5$4p	  &	$^1$S$  _{0}$	&   117.0469   &  117.0910   &   117.10551  \\
   60  &    2s$^2$2p$^5$4f	  &	$^3$D$  _{1}$	&   117.1009   &  117.1630   &   117.18462  \\
   61  &    2s$^2$2p$^5$4f	  &	$^1$G$  _{4}$	&  117.1128   & 117.1770   &   117.19678  \\
   62  &    2s$^2$2p$^5$4f	  &	$^3$D$  _{2}$	&  117.1216   & 117.1860   &   117.20638  \\
   63  &    2s$^2$2p$^5$4f	  &	$^3$G$  _{5}$	&  117.1226   & 117.1860   &   117.20569  \\
   64  &    2s$^2$2p$^5$4f	  &	$^3$F$  _{3}$	&  117.1618   & 117.2270   &   117.24811  \\
   65  &    2s$^2$2p$^5$4f	  &	$^1$D$  _{2}$	&  117.1651   & 117.2350   &   117.25364  \\
   66  &    2s$^2$2p$^5$4f	  &	$^1$F$  _{3}$	&  117.1727   & 117.2390   &   117.25952  \\
   67  &    2s$^2$2p$^5$4f	  &	$^3$F$  _{4}$	&  117.1862   & 117.2530   &   117.27332  \\
   68  &    2s$^2$2p$^5$4d	  &	$^3$F$^o_{2}$	&  118.3836   & 118.4350   &   118.45112  \\
   69  &    2s$^2$2p$^5$4d	  &	$^3$D$^o_{2}$	&  118.4268   & 118.4780   &   118.49398  \\
   70  &    2s$^2$2p$^5$4d	  &	$^1$F$^o_{3}$	&  118.4592   & 118.5090   &   118.52578  \\
   71  &    2s$^2$2p$^5$4d	  &	$^3$D$^o_{1}$	&  118.6287   & 118.6710   &   118.68965  \\
   72  &    2s$^2$2p$^5$4f	  &	$^3$G$  _{3}$	&  119.2006   & 119.2710   &   119.28929  \\
   73  &    2s$^2$2p$^5$4f	  &	$^3$F$  _{2}$	&  119.2205   & 119.2910   &   119.31213  \\
   74  &    2s$^2$2p$^5$4f	  &	$^3$G$  _{4}$	&  119.2203   & 119.2940   &   119.30882  \\
   75  &    2s$^2$2p$^5$4f	  &	$^3$D$  _{3}$	&  119.2266   & 119.2970   &   119.31618  \\
   76  &    2s$^2$2p$^5$5s	  &	$^3$P$^o_{2}$	&  126.1503   & 126.2080   &   126.22497  \\
   77  &    2s$^2$2p$^5$5s	  &	$^1$P$^o_{1}$	&  126.1823   & 126.2400   &   126.25746  \\
   78  &    2s2p$^6$4s  	  &	$^3$S$  _{1}$	&  126.5311   & 126.5660   &   126.57676  \\
   79  &    2s$^2$2p$^5$5p	  &	$^3$D$  _{2}$	&  126.8090   & 126.8570   &   126.88190  \\
   80  &    2s$^2$2p$^5$5p	  &	$^3$P$  _{1}$	&  126.8464   & 126.8640   &   126.91502  \\
   81  &    2s2p$^6$4s  	  &	$^1$S$  _{0}$	&  126.8375   & 126.8970   &   126.87346  \\
   82  &    2s$^2$2p$^5$5p	  &	$^3$D$  _{3}$	&  126.8857   & 126.9410   &   126.95811  \\
   83  &    2s$^2$2p$^5$5p	  &	$^3$P$  _{2}$	&  126.9340   & 126.9870   &   127.00426  \\
   84  &    2s$^2$2p$^5$5p	  &	$^1$P$  _{1}$	&  126.9367   & 126.9890   &   127.00670  \\
   85  &    2s$^2$2p$^5$5p	  &	$^1$S$  _{0}$	&  127.1699   & 127.2120   &   127.22363  \\
   86  &    2s$^2$2p$^5$5d	  &	$^3$P$^o_{0}$	&  127.5796   & 127.6410   &   127.65391  \\
   87  &    2s$^2$2p$^5$5d	  &	$^3$P$^o_{1}$	&  127.6080   & 127.6690   &   127.68237  \\
   88  &    2s$^2$2p$^5$5d	  &	$^3$F$^o_{4}$	&  127.6485   & 127.7090   &   127.72343  \\
   89  &    2s$^2$2p$^5$5d	  &	$^3$F$^o_{3}$	&  127.6517   & 127.7090   &   127.72533  \\
   90  &    2s$^2$2p$^5$5d	  &	$^3$P$^o_{2}$	&  127.6537   & 127.7130   &   127.72778  \\
\hline
\end{tabular}
\end{table}
\newpage
\begin{table}
\begin{tabular}{rllrrrrrrrrr} \hline
Index  &     Configuration        & Level               & GRASP        &   FAC1      &  FAC2 \\
\hline
   91  &    2s$^2$2p$^5$5d	  &	$^1$D$^o_{2}$	&  127.6843   & 127.7410   &   127.75723  \\
   92  &    2s$^2$2p$^5$5d	  &	$^3$D$^o_{3}$	&  127.7093   & 127.7650   &   127.78153  \\
   93  &    2s$^2$2p$^5$5d	  &	$^1$P$^o_{1}$	&  127.8377   & 127.8840   &   127.90177  \\
   94  &    2s$^2$2p$^5$5s	  &	$^3$P$^o_{0}$	&  127.8976   & 127.9260   &   127.93940  \\
   95  &    2s$^2$2p$^5$5s	  &	$^3$P$^o_{1}$	&  127.9415   & 127.9660   &   127.97990  \\
   96  &    2s$^2$2p$^5$5f	  &	$^3$D$  _{1}$	&  128.0211   & 128.0890   &   128.10779  \\
   97  &    2s$^2$2p$^5$5f	  &	$^3$D$  _{2}$	&  128.0352   & 128.1040   &   128.12248  \\
   98  &    2s$^2$2p$^5$5f	  &	$^3$G$  _{4}$	&  128.0408   & 128.1110   &   128.12852  \\
   99  &    2s$^2$2p$^5$5f	  &	$^3$G$  _{5}$	&  128.0448   & 128.1140   &   128.13205  \\
  100  &    2s$^2$2p$^5$5f	  &	$^3$D$  _{3}$	&  128.0595   & 128.1290   &   128.14743  \\
  101  &    2s$^2$2p$^5$5f	  &	$^1$D$  _{2}$	&  128.0683   & 128.1330   &   128.15776  \\
  102  &    2s$^2$2p$^5$5f	  &	$^1$F$  _{3}$	&  128.0692   & 128.1390   &   128.15764  \\
  103  &    2s$^2$2p$^5$5f	  &	$^3$F$  _{4}$	&  128.0770   & 128.1390   &   128.16560  \\
  104  &    2s2p$^6$4p  	  &	$^3$P$^o_{2}$	&  128.1320   & 128.1400   &   128.16658  \\
  105  &    2s2p$^6$4p  	  &	$^1$P$^o_{1}$	&  128.1372   & 128.1470   &   128.18768  \\
  106  &    2s2p$^6$4p  	  &	$^3$P$^o_{0}$	&  128.2730   & 128.1530   &   128.34535  \\
  107  &    2s2p$^6$4p  	  &	$^3$P$^o_{1}$	&  128.3123   & 128.1530   &   128.37384  \\
  108  &    2s$^2$2p$^5$5p	  &	$^3$D$  _{3}$	&  128.8614   & 128.1570   &   128.93661  \\
  109  &    2s$^2$2p$^5$5p	  &	$^3$P$  _{1}$	&  128.9718   & 128.1580   &   129.04802  \\
  110  &    2s$^2$2p$^5$5p	  &	$^1$D$  _{2}$	&  128.9715   & 128.1620   &   129.04755  \\
  111  &    2s$^2$2p$^5$5p	  &	$^3$P$  _{0}$	&  129.0372   & 128.1690   &   129.10388  \\
  112  &    2s2p$^6$4d  	  &	$^3$D$  _{1}$	&  129.6426   & 128.1730   &   129.66847  \\
  113  &    2s2p$^6$4d  	  &	$^3$D$  _{2}$	&  129.6433   & 128.1740   &   129.66991  \\
  114  &    2s2p$^6$4d		  &     $^3$D$  _{3}$   &  129.6544   & 128.3290   &   129.68199  \\
  115  &    2s$^2$2p$^5$5d	  &	$^3$F$^o_{2}$	&  129.7133   & 128.3570   &   129.79114  \\
  116  &    2s$^2$2p$^5$5d	  &	$^3$D$^o_{2}$	&  129.7363   & 128.9210   &   129.81430  \\
  117  &    2s$^2$2p$^5$5d	  &	$^1$F$^o_{3}$	&  129.7509   & 129.0320   &   129.82799  \\
  118  &    2s$^2$2p$^5$5d	  &	$^3$D$^o_{1}$	&  129.8364   & 129.0330   &   129.89680  \\
  119  &    2s2p$^6$4d		  &     $^1$D$  _{2}$   &  129.8254   & 129.0930   &   129.85765  \\
  120  &    2s$^2$2p$^5$5f	  &	$^3$G$  _{3}$	&  130.1195   & 129.6580   &   130.21107  \\
  121  &    2s$^2$2p$^5$5f	  &	$^3$G$  _{4}$	& 130.1302  & 129.6600   &   130.22156 \\
  122  &    2s$^2$2p$^5$5f	  &	$^3$F$  _{2}$	& 130.1389  & 129.6720   &   130.23026 \\
  123  &    2s$^2$2p$^5$5f	  &	$^3$F$  _{3}$	& 130.1457  & 129.7780   &   130.23506 \\
  124  &    2s2p$^6$4f		  &     $^3$F$^o_{2}$   & 130.4188  & 129.8010   &   130.47401 \\
  125  &    2s2p$^6$4f		  &     $^3$F$^o_{3}$   & 130.4215  & 129.8140   &   130.49580 \\
  126  &    2s2p$^6$4f		  &     $^3$F$^o_{4}$   & 130.4331  & 129.8430   &   130.54657 \\
  127  &    2s2p$^6$4f		  &     $^1$F$^o_{3}$   & 130.4497  & 129.8830   &   130.54456 \\
\hline				  				        				   
\end{tabular}	
									            			    
\begin {flushleft}									            			    
\begin{tabbing} 									      	
aaaaaaaaaaaaaaaaaaaaaaaaaaaaaaaaaaaa\= \kill						      	       
GRASP: Earlier results of Singh and Aggarwal \cite{sa}  with the {\sc grasp} code  \\			
FAC1: Earlier results of Singh and Aggarwal \cite{sa}  with the {\sc fac} code  \\ 			
FAC2: Present results  with the {\sc fac} code  			      		      	
\end{tabbing}										      	
\end {flushleft}								
\end{table}


\begin{thebibliography}{999}
\bibitem{sa} N Singh and S Aggarwal,  {\em Pramana --  J. Phys.}  {\bf 89}, 79 (2017)
\bibitem{brlike} K M Aggarwal and F P  Keenan, {\em Phys. Scr.} {\bf 89}, 125404 (2014)
\bibitem{w66a} K M  Aggarwal,  {\em Chin. Phys. B} {\bf 25},  043201 (2016)
\bibitem{w66b} K M  Aggarwal, {\em Atoms} {\bf 4}, 4030024 (2016)
\bibitem{hlm} A Hibbert, M Le Dourneuf and M Mohan, {\em At. Data Nucl. Data Tables} {\bf 53}, 23 (1993)
\bibitem{jqs} K M  Aggarwal,  {\em J. Quant. Spectros. Rad. Transfer} {\bf 166},  108 (2015)
\bibitem{kma} K M  Aggarwal, {\em Atoms} {\bf 5}, 5040037 (2017)
\bibitem{nelike} K.M. Aggarwal. At. Data Nucl. Data Tables {\bf 124}, in press (2018) --  http://arxiv.org/abs/1803.04203


\end{thebibliography}
\end{document}